# Antisymmetric magnetoresistance in van der Waals $Fe_3GeTe_2$/graphite/$Fe_3GeTe_2$ tri-layer heterostructures

Sultan Albarakati[1, †], Cheng Tan[1, †], Zhong-Jia Chen[2, †], James G. Partridge[1], Guolin Zheng[1], Lawrence Farrar[1], Edwin L.H. Mayes[1], Matthew R. Field[1], Changgu Lee[3], Yihao Wang[4], Yiming Xiong[4], Mingliang Tian[4], Feixiang Xiang[5], Alex R. Hamilton[5], Oleg A. Tretiakov[5,6], Dimitrie Culcer [5,*] Yu-Jun Zhao[2,*] and Lan Wang[1,*]

[1]School of Science, RMIT University, Melbourne, VIC 3000, Australia.

[2]Department of Physics, South China University of Technology, Guangzhou, Guangdong 510640, China

[3]Center for Quantum Materials and Superconductivity (CQMS) and Department of Physics, Sungkyunkwan University, Suwon, Republic of Korea.

[4]Anhui Province Key Laboratory of Condensed Matter Physics at Extreme Conditions, High Magnetic Field Laboratory of the Chinese Academy of Sciences, Hefei, Anhui 230031, China.

[5] School of Physics and ARC Centre of Excellence in Future Low-Energy Electronics Technologies, UNSW Node, University of New South Wales, Sydney NSW 2052, Australia.

[6] Institute for Materials Research, Tohoku University, Sendai 980-8577, Japan.

† These authors contributed equally to the paper.

[*] Corresponding authors. Correspondence and requests for materials should be addressed to Lan Wang:

lan.wang@rmit.edu.au, and Yu-Jun Zhao: zhaoyj@scut.edu.cn and Dimitrie Culcer: d.culcer@unsw.edu.au

## Abstract

Van der Waals (vdW) ferromagnetic materials are rapidly establishing themselves as effective building blocks for next generation spintronic devices. When layered with non-magnetic vdW materials, such as graphene and/or topological insulators, vdW heterostructures can be assembled (with no requirement for lattice matching) to provide otherwise unattainable device structures and functionalities. We report a hitherto rarely seen antisymmetric magnetoresistance (MR) effect in van der Waals heterostructured $Fe_3GeTe_2$/graphite/$Fe_3GeTe_2$ devices. Unlike conventional giant magnetoresistance (GMR) which is characterized by two resistance states, the MR in these vdW heterostructures features distinct high, intermediate and low resistance states. This unique characteristic is suggestive of underlying physical mechanisms that differ from those observed before. After theoretical calculations, the three resistance behavior was attributed to a spin momentum locking induced spin polarized current at the graphite/FGT interface. Our work reveals that ferromagnetic heterostructures assembled from vdW materials can exhibit substantially different properties to those exhibited by similar heterostructures grown in vacuum. Hence, it highlights the potential for new physics and new spintronic applications to be discovered using vdW heterostructures.

## MAIN TEXT

## Introduction

Microelectronic devices, such as diodes and transistors, can be incorporated into large integrated structures capable of performing diverse tasks including logical operations (i.e. computing devices) and high-density information storage. Despite a vast range of architectures and applications, the fundamental principle of operation for all devices relies on the manipulation of a single quantity – the charge of the electron. The concept of utilizing spin in addition to charge in microelectronics is termed spintronics. Giant magnetoresistance (GMR), which progressed from discovery to applications (modern disk storage) within a decade, perfectly illustrates the impact of spintronics. By manipulating the electron charge and spin simultaneously, ultra-high speed and low-power electronic devices with enhanced functionality can be realized. As a result, spintronics is now one of the most important technological research fields, encompassing electronics, materials science and condensed matter physics[1,2].

With the emergence of two dimensional (2D) materials, van der Waals (vdW) ferromagnetic materials have attracted great interest. 2D ferromagnetism has been confirmed in three monolayer vdW materials, $CrI_3$ (insulator), $Cr_2Ge_2Te_6$ (insulator) and $Fe_3GeTe_2$ (metal)[(3-6)]. So far, various spintronic devices have been fabricated based on these materials[(7-15)]. vdW ferromagnetic metal $Fe_3GeTe_2$ (FGT) has been studied for many years[(16-30)]. Only very recently, it was realized that nanoflake vdW FGT is a promising vdW ferromagnetic metal for spintronics with a near-square shaped hysteresis loop, large coercivity and perpendicular magnetic anisotropy[(16)], making it a perfect vdW ferromagnetic metal for spintronic devices[(9)] and fundamental spintronic research. Furthermore, materials with very different crystal structures and lattice constants can be layered in vdW heterostructures without the deleterious effects seen in conventional thin film heterostructures. Hence, a major design restriction is removed to provide near limitless possibilities for novel spintronic device architectures.

The discovery of GMR[(31,32)] was honored with the Nobel Prize in 2007 and introduced the second fundamental property of the electron - its spin, into microelectronics. A standard GMR device has a tri-layer structure with two ferromagnetic metals separated by a non-ferromagnetic metal. To date, most GMR devices have been fabricated from metallic thin films grown in a high vacuum chamber. The non-ferromagnetic layers in these GMR devices are typically metals, such as Cr[(32)], Cu[(33)] or Pd[(34)], but interest is turning to non-ferromagnetic conductive materials with desirable properties, such as multilayer graphene[(35,36)]. Incompatibility with magnetic sputtering (commonly used for growth of GMR structures) is a concern with multilayer graphene but once again, this is not a limitation in vdW heterostructure devices. Non-magnetic conductors such as graphene can be sandwiched between vdW ferromagnets with atomically flat and ultraclean interfaces.

We have fabricated FGT/graphite/FGT devices using a pick-up transfer technique[(37,38)] and investigated their magnetoresistance (MR) behavior, with surprising results. The thickness of graphite (multilayer graphene) within the structure was varied from 3 nm to 11 nm. In a standard GMR experiment, a tri-layer device shows a symmetric MR effect. When the magnetic moments in the two ferromagnetic layers point in opposite directions, the tri-layer structure adopts a high resistance state. When the magnetic moments point in the same direction, the tri-layer GMR device adopts a low resistance state. The FGT/graphite/FGT devices studied here display an intermediate resistance state when the magnetic moments are parallel. When the magnetic moments in the two FGT layers are antiparallel, the devices exhibit high resistance for positive magnetic fields and low resistance for negative magnetic fields (at times inversely). We propose that the interfaces between the graphite flake and the FGT flakes and the strong spin orbit coupling (SOC) in FGT, discussed below, are responsible for the observed three state MR behavior.

## Results

In our experiment, eleven FGT/graphite/FGT devices with differing thicknesses of graphite layers were fabricated. The labels and dimensions are shown in the supplementary information. Fig. 1A shows the optical and atomic force microscopy (AFM) image of a trilayer heterostructure (sample FPC3) with a graphite layer sandwiched by two FGT layers. In the AFM image, the blue section defines the top FGT layer, the red section defines the graphite layer and the yellow section defines the lower FGT layer. The device current flowed in plane from the source to the drain and the longitudinal resistance, $R_{xx}$ and the anomalous Hall resistance, $R_{xy}$ were measured. As FGT shows strong perpendicular anisotropy[(16)], the magnetic field applied perpendicular to the device surface was swept from 1 T to -1 T and then from -1 T to 1 T. The diagram in fig. 1B illustrates the normal (two-state) GMR effect, which is symmetric with respect to the applied magnetic field. In contrast, the FGT/graphite/FGT device (sample FPC3) displays an antisymmetric MR effect, as shown in fig. 1C. The anomalous Hall measurements of the FGT/graphite/FGT device show very sharp magnetic transitions, in agreement with our previous work[(16)]. Namely, an individual FGT nanoflake shows a near-square shaped magnetic loop and has a single domain structure in the regime away from coercivity. At the positive value of the saturation magnetic field, the magnetic moments of both FGT layers point along the positive direction. When the magnetic field reaches a certain regime of negative values, the magnetic moments of the two FGT layers point in opposite directions due to the different coercivities of these layers. The MR exhibits a plateau with a sharp transition while the anomalous Hall resistance exhibits step-like behavior. When the magnetic field reaches its negative saturation value, the moments in both FGT layers point along the negative direction. $R_{xx}$ in this configuration is equal to $R_{xx}$ when the moments in both layers point in the positive direction. When the field is swept back from the saturation value to a certain regime of positive values, the magnetic moments in the two FGT layers point in opposite directions. Again, $R_{xx}$ exhibits a plateau and $R_{xy}$ is stepped. However, unlike conventional GMR with two resistance states, the FGT/graphite/FGT shows three resistance states, namely high resistance (antiparallel magnetic moments), intermediate resistance (parallel magnetic moments), and low resistance (antiparallel magnetic moments). The magnitudes of the MR effect at 50 K in all eleven devices (with differing graphite layer thicknesses) are shown in fig. 1D. These results show that the antisymmetric MR effect is independent of the thickness of the graphite layer. It is well known that graphene and nanoflake graphite display very large MR. However, the FGT/graphite/FGT devices show very small MR at 1 and -1 Tesla, which demonstrates the clean FGT/graphite interface. If the FGT/graphite interface is not clean (organic residues or bubbles), $R_{xx}$ is characterized by a large quasi-linear MR (Fig. S3c and Fig. S3d). This is because current mainly flows in the graphite layer and consequently, charge carriers are not scattered by the magnetic moments in the FGT layers.

Additional experimental work was performed to establish whether the observed phenomena were intrinsically generated from the heterostructure. Firstly, it was a concern that the asymmetric nanoflakes within the devices could cause the anti-symmetric $R_{xx}$ due to the anomalous Hall effect. Secondly, the number of states in FGT/graphite/FGT heterostructure has not been confirmed. In fig. 1C, a small "loop" (surrounded by a dark blue dashed line, open in vertical direction) between the 2 peaks can be observed, which may originate from the Hall resistance or indicate a 4-state MR effect.

To address the two issues, three symmetric devices were prepared using focused ion beam (FIB) etching.  Fig. S4a shows the image of one etched device (sample FPC5). Alongside the device in fig. S4 are the $R_{xx}$ and $R_{xy}$ curves measured at 50 K. To elucidate the effects of device symmetry, we performed the FIB etching in two steps. In the first step, the parts identified by the orange dashed line were etched. Compared with the original $R_{xx}$ (Fig. S4b), the "loop" (surrounded by a green dash line) in $R_{xx}$ (B) between the two peaks decreases after this first etch step (Fig. S4c). In the second FIB etch step, the parts identified by the red dashed line were etched and afterwards, only the parts surrounded by the blue dashed line remained. Fig. S4d shows the $R_{xx}$ and $R_{xy}$ curves of the sample at 50 K after the second etch. The $R_{xy}$ curve indicates that the completed device is symmetric, while the $R_{xx}$ curve still shows the antisymmetric MR effect and the "loops" between the peaks disappears. These observations prove that the phenomenon is not an artifact caused by asymmetry in the devices and/or contacts but a genuine three-resistance state system originating from the FGT/graphite/FGT heterostructures. During this study we found that despite precautionary measures, FIB etching could cause damage to the top FGT layer and affect its sharp magnetic transition. We therefore chose to remove FIB from the fabrication scheme to ensure that the MR effect was observed in devices composed of pristine FGT layers.

Fig. 2 shows the temperature dependence of the antisymmetric MR with the magnetic field applied perpendicular to a device (sample FPC1) plane. The magnitude of the antisymmetric MR decreases with increasing temperature and disappears when $T > 140$ K. As shown in our previous work, the magnetic properties of FGT nanoflakes undergo a sharp transition near 150 K. Below 150 K, they exhibit a near-square shaped loop, while the magnetic remanence and coercivity decrease sharply to near zero when $T > 150$ K, although the Curie temperature is near 200 K. Taken together the aforementioned results indicate a magnetic hysteresis loop with nonzero remanence is essential for the antisymmetric MR observed in our FGT/graphite/FGT devices.

Fig. 3 shows the angular dependence of the antisymmetric MR effect at 20 K. The angles shown in the figure represent the angles between the magnetic field and the direction perpendicular to the plane of the device. As shown in fig. 3A, the magnitude of the MR effect does not change as the angle is varied. However, the two antisymmetric MR plateaus flip in the region between θ = 70° to 72°. From θ = 0 to 70°, the high resistance state appears at positive fields and the low resistance state appears at negative fields. The width of the resistance plateau decreases with increasing angle θ. When θ exceeds 72°, the resistance plateaus flip: the high resistance state appears at negative fields and the low resistance state appears at positive fields. With a further increase in θ, the resistance plateaus increase in width. At θ = 85°, the resistance plateau shows several smaller plateaus. To understand this behavior, we compared the $R_{xx}$(B) and $R_{xy}$ (B) curves (Fig. 3A and 3B). As the thickness of the top FGT layer and bottom FGT layer are 21.7 nm and 42 nm, respectively, the decrease in $R_{xy}$ is less if a flip occurs in the top FGT layer than if a flip occurs in the bottom FGT layer. Hence, the layer that flips is clearly distinguishable. From 0° to 70°, the $R_{xy}$ curve (Fig. 3B) shows a smaller decrease first and a larger decrease after the resistance plateau when the magnetic field sweeps from 1 T to -1 T, showing that the thinner top FGT layer is first to flip. At the location of the plateau in $R_{xx}$, the moments in both layers point into the graphite layer – this will be referred to as the “IN” configuration. When the magnetic field is scanned back from -1 T to 1 T, the first layer to flip is again the top layer with a smaller decrease in $R_{xy}$. However, the magnetic moments are now in the “OUT” configuration. Namely, the magnetic moments in both layers point out of the tri-layer structure. Therefore, from θ = 0° to θ = 70°, the “IN” configuration (at negative field) shows the low resistance state while the “OUT” configuration (at positive field) shows the high resistance state. When θ > 70°, the states flip and the high resistance state appears at negative field, while the low resistance state appears at positive field. Further investigation of the $R_{xy}$ indicates that the magnetic moment of the bottom layer flips first when θ > 70°, which is opposite to the condition when θ < 70°. Hence, the low resistance state still corresponds to the “IN” state and the high resistance state still corresponds to the “OUT” state. Hence, from the analysis of the angular dependent $R_{xx}$ and $R_{xy}$ measurements, we conclude that the magnetic configuration determines the resistance states; When the magnetic moments in the two FGT layers are parallel, the FGT/graphite/FGT device is in the intermediate resistance state and when they are antiparallel, the “IN” state causes low resistance and the “OUT” state causes high resistance.

To further understand the three-state MR effect, we performed a series of measurements with opposing current directions and flipped device orientation. In the current direction dependent MR measurements, the applied magnetic field is perpendicular to the sample surface ($\theta = 0^{\circ}$). As shown in fig. 4A, the "IN" state shows the low and high resistance state with the positive and negative current, respectively, while the "OUT" state shows the low and high resistance state with the negative and positive states, respectively. In the device orientation dependent MR experiments, the MR was first measured with the FGT top-layer facing upwards. Thereafter, the device was inverted (with the original top FGT at the bottom of the device) and the MR was remeasured. As shown in fig. 4B, the corresponding high and low resistance states reverse as well. We also measured the current density dependent plateau resistance. As shown in fig. 4C, the antisymmetric MR effect is independent of the current density.

Antisymmetric MR is uncommon but it has been reported to have occurred in two different systems. The first was composed of single layer magnetic thin films with perpendicular anisotropy[(39,40)]. In a magnetic thin film with perpendicular anisotropy, two single magnetic domains separated by a $180^{o}$ domain wall are formed. The magnetic moments in the two single domains point 'up' and 'down', respectively. When a current is passed perpendicularly through the domain wall, a perpendicular electric field can form in the vicinity of the domain wall due to the anomalous Hall effect. An antisymmetric MR can then be observed if the magnetic domain wall is driven by an applied perpendicular magnetic field. The second system reportedly produced antisymmetric MR was a novel topological materials-based magnetic heterostructure composed of $CrSb/(Bi,Sb)_2Te_3/CrSb$ layers grown by molecular-beam epitaxy[(41)]. In this very recent report, the antisymmetric MR is proposed to be related to quantum anomalous Hall effect. As shown in Fig 4A and 4B, the high and low resistance states switch to the low and high resistance states, respectively when the direction of current flow is reversed. This phenomenon indicates the strong correlation between the electron spin and momentum, namely spin-momentum locking. There are two possible origins of spin momentum locking, topological surface states on topological materials[(42)] and an SOC induced Rashba-split two-dimensional electron gas[(43)]. Recent work has identified FGT as a ferromagnetic topological nodal line semimetal[(24)], which is expected to have a strong spin-orbit interaction. To investigate the origin of spin momentum locking in the FGT/graphite/FGT devices, we performed a series of density functional theory (DFT) calculations. The results of these calculations demonstrate that the spin of the surface state of FGT (the Fermi arc) points in the same direction as the magnetization (details in supplementary section S8). Therefore, the Fermi arc does not show spin momentum locking. Thus the spin momentum locking can only originate from the SOC induced Rashba-split two-dimensional electron gas. In agreement with a recent report[(24)], our calculation also shows SOC induced band splitting in FGT. This substantiates the assertion that an SOC induced Rashba-split two-dimensional electron gas causes spin momentum locking. Reversal of the current direction then flips the spins of the transported electrons.

Based on the current dependent spin orientation at the FGT surface, we propose a tentative model that elucidates the appearance of a three-state MR in a FGT/graphite/FGT heterostructure. In an external electric field the SOC induced spin momentum locking will generate a sizable spin current transverse to the original electric current on the FGT surface (note that this is not a spin-Hall current[(44)]). Although the resistivity of graphite is $10^4$ orders smaller than that of FGT, these spin currents will still be flowing through the two FGT/graphite interfaces. Because the thicknesses of our graphite layers, which are 2~11 nm, are much smaller than the electron mean free path in graphite[(45)], the electrons transported through the graphite can enter the FGT layers and scatter within them. This picture is well supported by the $R_{xy}$ measurements. If the electrons only flowed in the graphite layer, the $R_{xy}$ would not display any anomalous Hall effect. This is in contradiction to the experimental results. In this scenario, the current direction determined electron spin polarization of the spin current may coincide with or be opposite to the direction of the magnetization in one FGT layer. If the two directions are antiparallel, this will lead to larger scattering and consequently higher resistance than observed when the polarizations of the spin current and the magnetization are parallel. With this in mind one can distinguish three situations: (I) at both interfaces, the spin polarization and magnetization are parallel which leads to the lowest observed resistance, (II) at both interfaces the spin polarization and magnetization are antiparallel, which leads to largest observed resistance, and (III) at one of the interfaces the spin polarization and magnetization are parallel, but at the other they are antiparallel, which leads to an intermediate resistance.

The above theoretical description explains all the experimental results, as shown in fig. S9. Fig. S9a illustrates the SOC induced spin momentum locking in the Rashba spin splitting surface states. When the current flows on the surface, spin current is generated due to spin momentum locking. As mentioned above, the current mainly flows in the graphite layer and at the interfaces. For the top and bottom FGT layer, the current flows at their bottom and top surface, respectively. Hence, the spin momentum locking induced current shows opposite spin orientations at the two interfaces. In this case, when the magnetizations of the two FGT layers point in the same direction, situation

III (intermediate resistance state) as mentioned above occurs, which is shown in fig. S9b and fig. S9c. When the magnetizations of the two FGT layers point in opposite directions, the situation I (low resistance state) and situation II (high resistance state) occur, which are plotted in fig S9d and fig. S9e, respectively. If the direction of current flow is reversed, the spin polarization of the electrons at the interface will be switched but the magnetization in the FGT layers is maintained, hence the intermediate state remains, while the low and high resistance states switch. If, on the other hand, the direction of current flow is maintained and the device is flipped upside down, then the magnetization configuration switches while the electron spin directions at the interface are maintained. Hence, again, the intermediate state remains unchanged, while the low and high resistance states switch. Finally, as shown in fig 3 and fig S5a, the magnetization configuration changes with the tilt-angle and temperature, which determines whether the high and low resistance states appear at negative or positive field. As the magnetic dynamics of FGT layers are different for different devices, the evolution of the antisymmetric MR with tilt angle and temperature also varies with devices (Details in supplementary section 9).

## Discussion

In summary, we have observed a previously unreported three-state MR effect in FGT/graphite/FGT vdW heterostructure devices. This surprising result is contrary to previously held beliefs regarding GMR and its mechanisms. Band structure calculations performed on FGT have enabled us to propose that this three-state MR arises due to spin momentum locking at the FGT/graphite interface. This work highlights the potential to discover entirely new phenomena and applications in vdW heterostructure based spintronic devices.

## Materials and Methods

### Single crystal growth

The FGT single crystals for the top and bottom device layers were grown by different methods to achieve different coercive fields. The FGT single crystals from C.L.'s group were grown by the chemical vapor transport (CVT) method. Iodine (5 mg/cm$^2$) served as a transport agent was mixed with Fe, Ge and Te powder (3:1:5). The mixed powder was sealed into an evacuated quartz glass ampoule, which was then placed in a tubular furnace. The temperature of the furnace was ramped to 700 ºC at a rate of 1 ºC per minute and was maintained at the set point for 96 hours. The ampoule was subsequently cooled to 450 ºC for 250 hrs. This slow cooling promotes high crystallinity. FGT single crystals from Y.X.'s group were also grown by the chemical vapor transport method but the transport agent $I_2$ was replaced by $TeCl_4$ to achieve larger single crystals suitable for transport measurements.

**Device fabrication**

The FGT and few-layer graphite flakes were mechanically exfoliated onto $SiO_2$/Si wafers in a glove box with oxygen and water levels below 0.1 ppm. The nanoflakes were examined by optical microscopy. Atomically smooth flakes were identified with thickness of 5-20 nm for the top FGT, 40~80 nm for the bottom FGT and 2-15 nm for the graphene. All three layers were manipulated using a polymer-based dry transfer technique. The polymer was dissolved in chloroform outside of the glove box for approximately one minute and the sample was spin-coated immediately after the dissolution to minimize the exposure to ambient conditions. Cr/Au (5/120 nm) metal electrodes were deposited onto the overlap area by electron beam evaporation following a standard electron beam lithography technique using poly methylmethacrylate (PMMA) 950 7A as the electron-beam resist (lift-off) layer.

**FIB etch**

The FIB etching of the heterostructure device was performed using an FEI Scios DualBeam system. The Ga ion energy was 30 keV and the beam current was 1.5 pA. The duration of FIB etching was limited to 10 s and with a beam-blocked interval of at least 10 s to minimize heating of the sample. The FIB was not scanned over the device region, which is the part surrounded by the blue dashed line in fig. S4a.

**Electrical measurement**

The transport measurements were performed in a Physical Property Measurement System (Ever Cool II, Quantum Design, San Diego, CA, USA) with a base temperature of 1.8 K and with magnetic field up to 9 T. The devices were mounted on a Horizontal Rotator probe, which allows device rotations around an axis perpendicular to the magnetic field.

**Acknowledgments**：**Funding**：This research was supported by the Australian Research Council Centre of Excellence in Future Low-Energy Electronics Technologies (CE170100039), Natural Science Foundation of China (11574088), National Key Research and Development Program of China (2016YFA0300404) and the Institute for Information & Communications Technology Promotion (IITP) grant (B0117-16-1003, Fundamental technologies of 2D materials and devices for the platform of new-functional smart devices). Author Contributions: L.W. conceived and designed the research. C.L., Y.W., Y.X., and M.T. synthesized the material, S.A. and C.T. fabricated the heterostructures. S.A., C.T. and G.Z did the device fabrication. S.A., C.T., F.X., A.H., and L.W. performed the electron transport measurements, S.A., C.T., A.H., O.A.T., D.C. and L.W. did the data analysis and modeling. Z.C. and Y.Z. did the band structure calculation. L.F., E.L.H.M., M.R.F. performed the TEM scan for the cross-section of heterostructures. S.A., C.T., Z.C., Y.Z, J.P., D.C. and L.W. wrote the paper with the help from all of the other co-authors. **Competing interests:** The authors declare that they have no competing interests. **Data and materials availability:** All data needed to evaluate the conclusions in the paper are present in the paper and/or the supplementary materials. Additional data related to this paper may be requested from the authors.

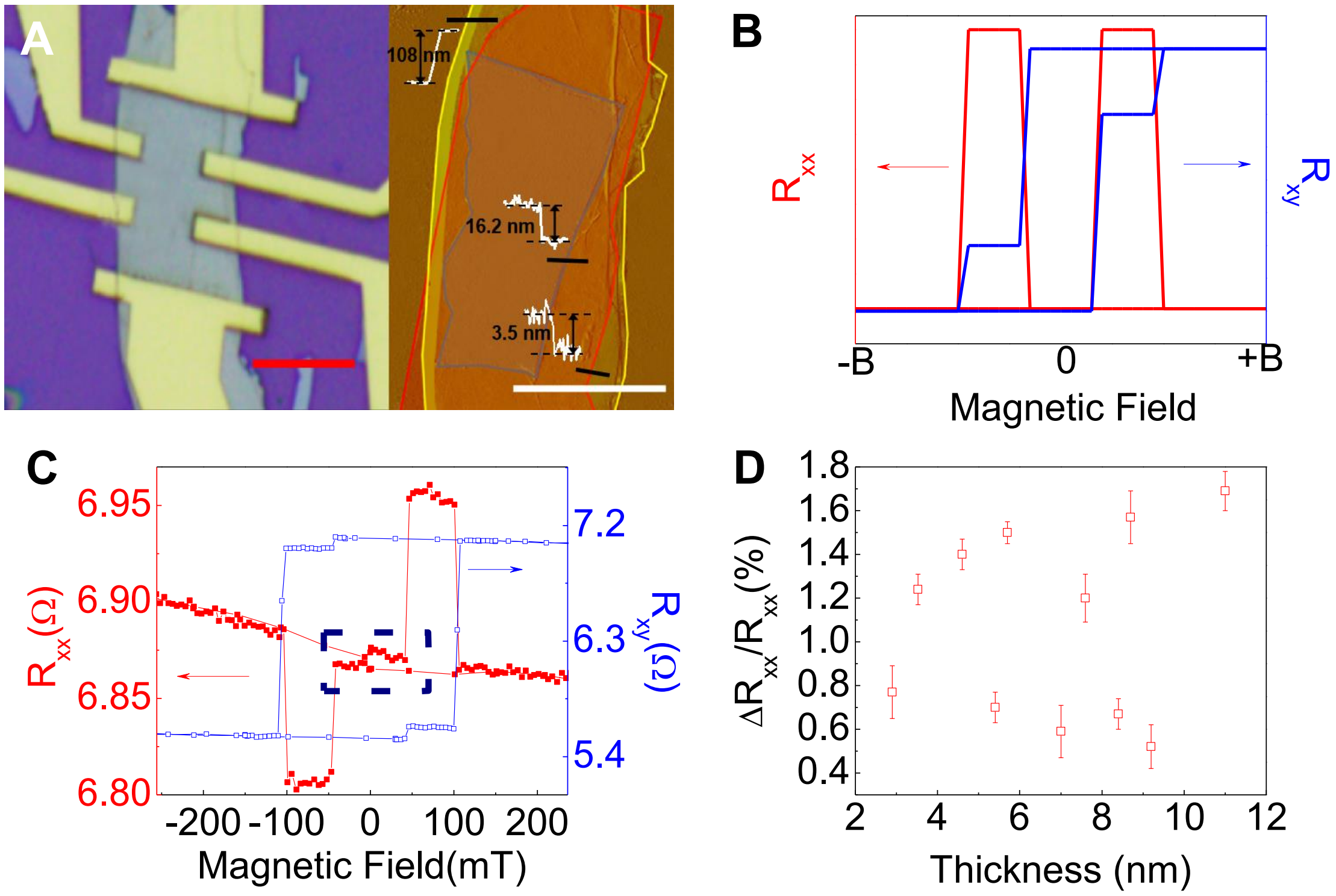

**Fig. 1 Overview of the MR effect in FGT/graphite/FGT heterostructures.** (**A**) Optical and AFM images of a FGT/graphite/FGT heterostructure. The device number is FPC3. Both scale bars represent 5 μm. The regions surrounded by the blue line, red line and yellow line represent the top FGT layer, graphite layer and bottom FGT layer, respectively. (**B**) A schematic diagram for the transport behavior of a typical GMR effect. (**C**) Field dependent $R_{xx}$ and $R_{xy}$ measurements of an FGT/graphite/FGT heterostructure (sample FPC3) at 50 K. A "loop" surrounded by a dark blue dashed line is shown in the $R_{xx}(B)$ curve. (**D**) $\Delta R_{xx}/R_{xx}$ values for samples with various thicknesses of graphite layer. All the data are calculated for measurements at 50 K. The error bars come from the noise of the measurement.

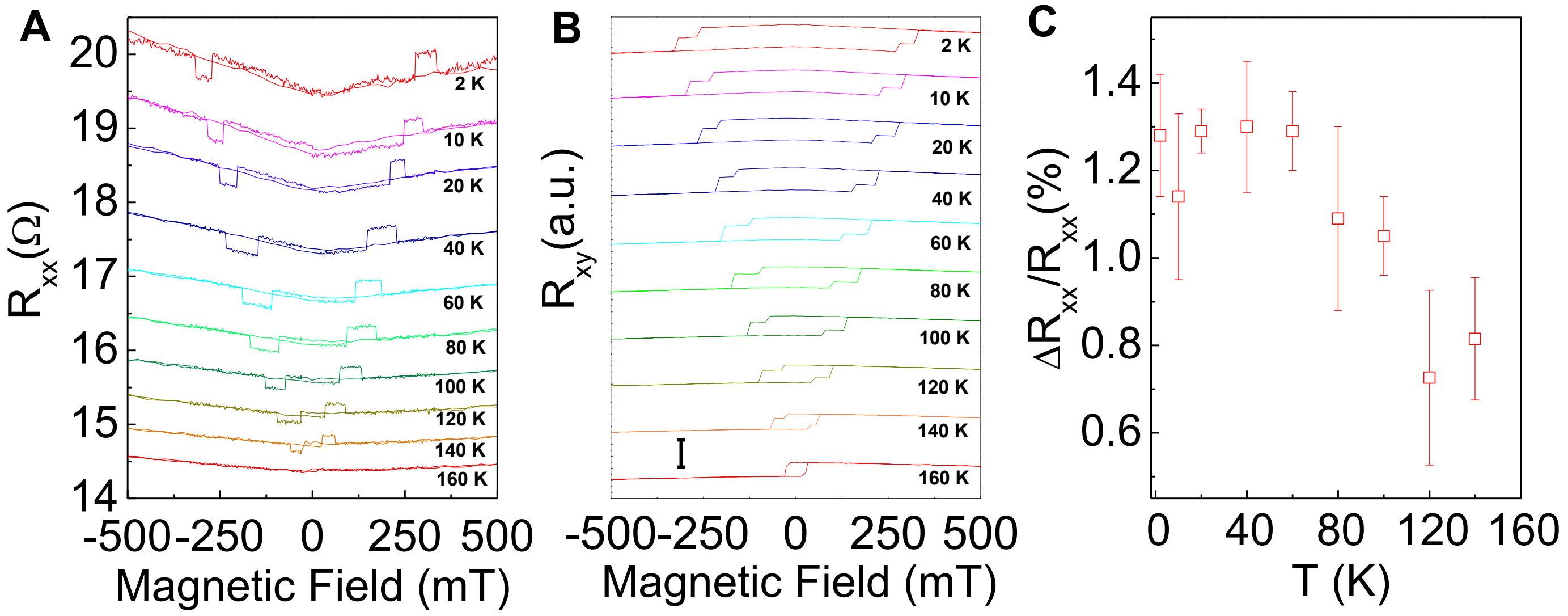

**Fig. 2 Temperature dependent transport measurement for sample FPC1.** (**A**) The $R_{xx}$ curves in a FGT/graphite/FGT device at different temperatures. (**B**) Corresponding $R_{xy}(B)$ curves at different temperatures. The scale bar represents 3 Ω. (**C**) Temperature dependence of $\Delta R_{xx}/R_{xx}$ values. The error bars are defined by the noise level.

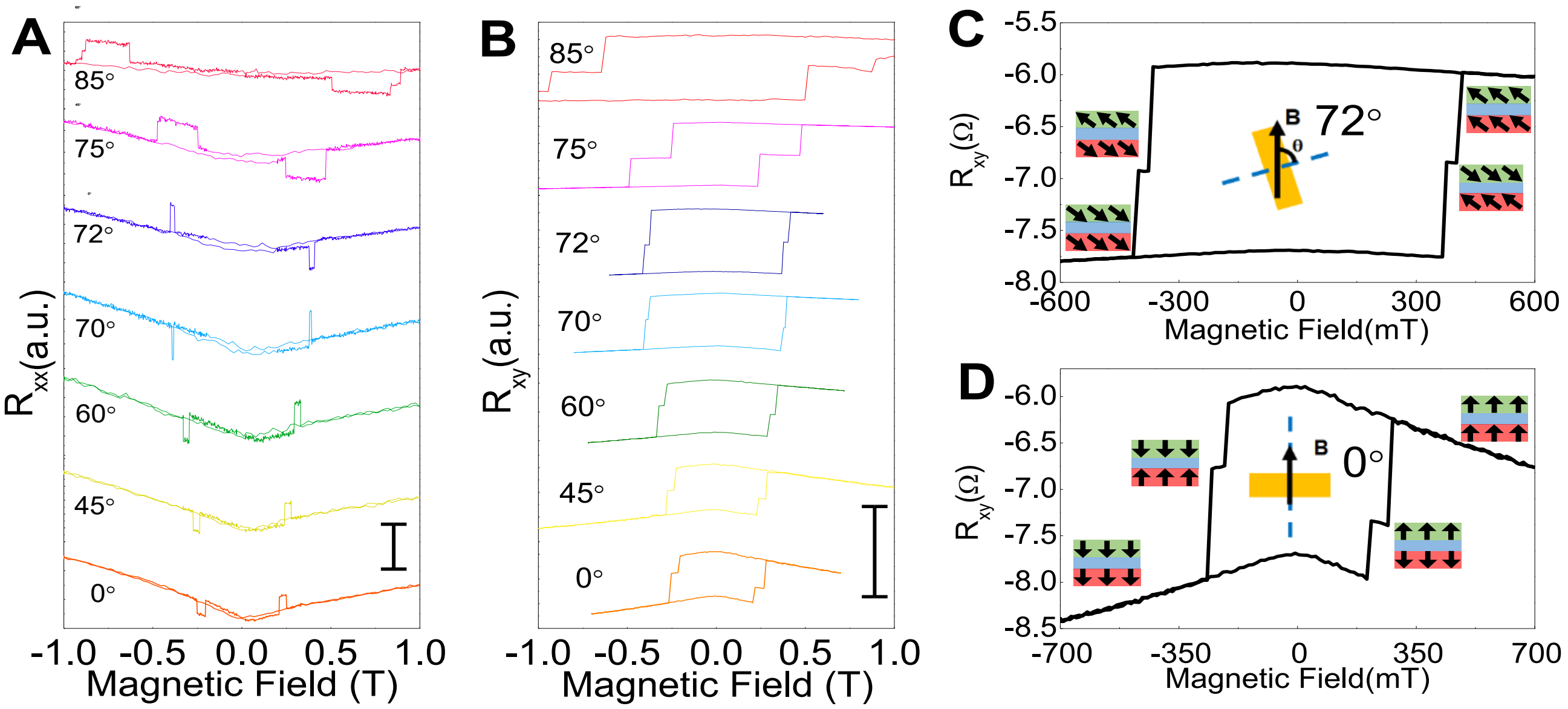

**Fig. 3 Angular dependent transport measurements for sample FPC1 at 20 K.** (**A**) The $R_{xx}(B)$ curves of a FGT/graphite/FGT device at different tilt angles at 20 K. 0° means the magnetic field perpendicular to the sample surface. The scale bar represents 1.5 Ω. (**B**) Corresponding $R_{xy}(B)$ curves at different tilt angles at 20 K. The scale bar represents 4.5 Ω. (**C**) $R_{xy}(B)$ curve at 72°, the magnetic moments in the bottom layer flips first with increasing magnetic field. (**D**) $R_{xy}(B)$ curve at 0°, the magnetic moments in the top layer flips first with increasing magnetic field.

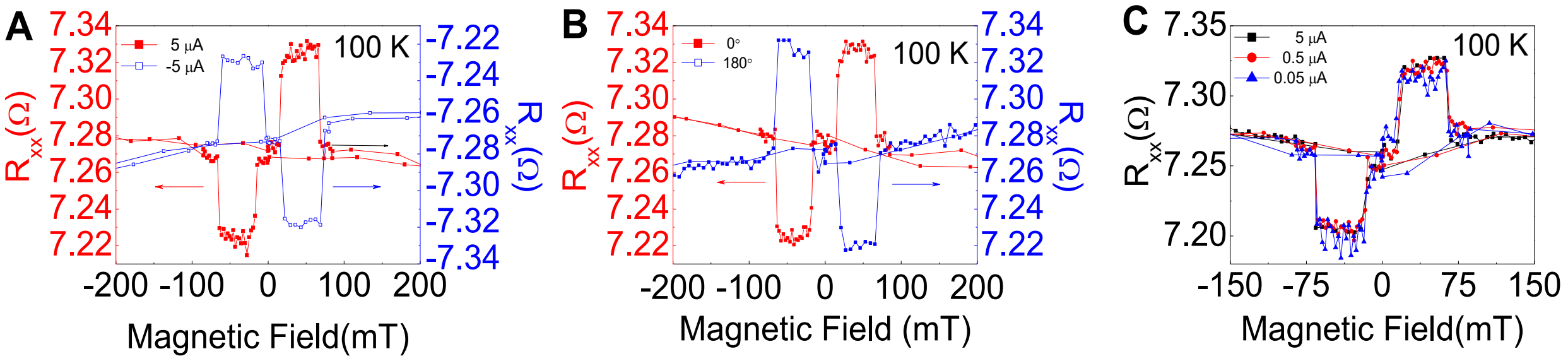

**Fig. 4 Current dependent transport measurements on FPC3 at 100 K.** (**A**) $R_{xx}$(B) with different current directions. (**B**) $R_{xx}$ measured when the top FGT layer faces upwards(0º) and downwards(180º). (**C**) $R_{xx}$(B) with different magnitudes of current.

Supplementary Materials for

# Antisymmetric magnetoresistance in van der Waals $Fe_3GeTe_2$/graphite/$Fe_3GeTe_2$ tri-layer heterostructures

Sultan Albarakati[1, †], Cheng Tan[1, †], Zhong-Jia Chen[2, †], James G. Partridge[1], Guolin Zheng[1], Lawrence Farrar[1], Edwin L.H. Mayes[1], Matthew R. Field[1], Changgu Lee[3], Yihao Wang[4], Yiming Xiong[4], Mingliang Tian[4], Feixiang Xiang[5], Alex R. Hamilton[5], Oleg A. Tretiakov[5,6], Dimitrie Culcer [5,*] Yu-Jun Zhao[2,*] & Lan Wang[1,*]

[*] Corresponding authors: Lan Wang: lan.wang@rmit.edu.au , Yu-Jun Zhao: zhaoyj@scut.edu.cn and Dimitrie Culcer: d.culcer@unsw.edu.au

## Supplementary Information Outline:

## 1. Ohmic contacts

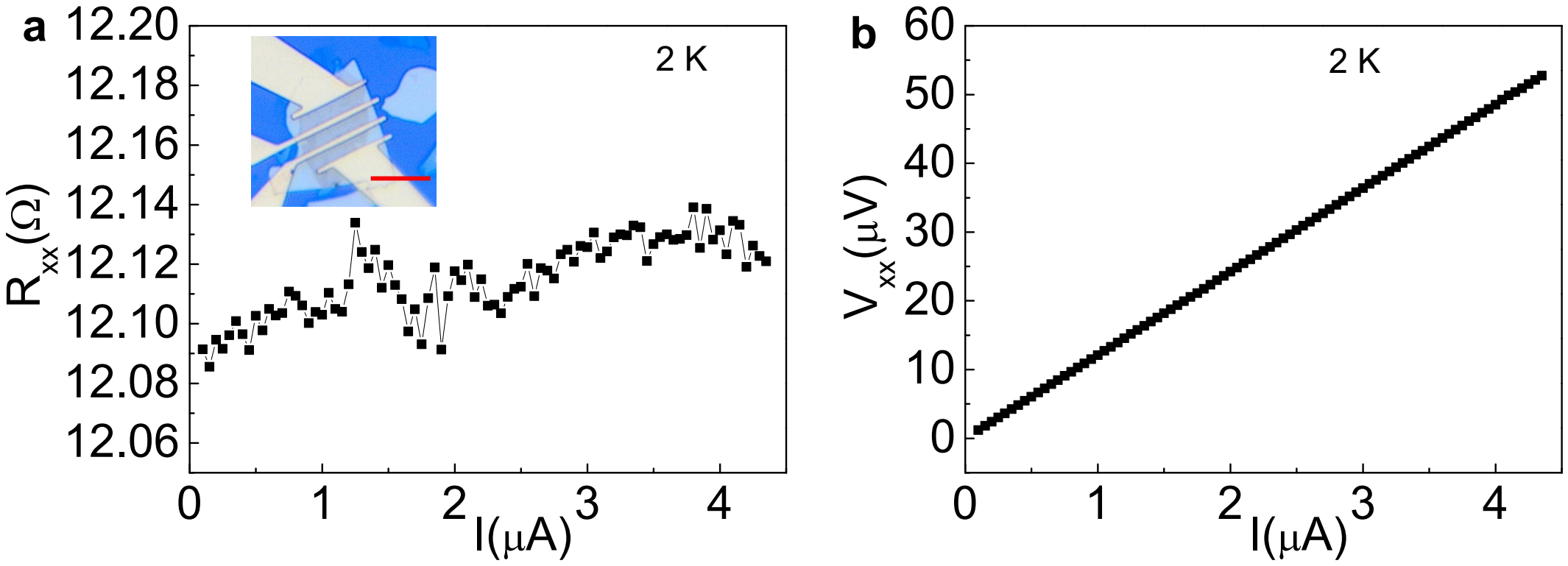

**Fig. S1 Confirmation of ohmic device contacts.** **(a)** $R_{xx}$ vs current curve at 2 K. Inset is an image of the device under test with a scale bar representing 10 μm. **(b)** The corresponding I-V curve derived from **(a).**

## 2. Dimensions of all FGT/Graphite/FGT devices

| Device label | Upper FGT(nm) | Lower FGT(nm) | Graphite (nm) | Length × Width (μm×μm) |
|---|---|---|---|---|
| **FPC1** | 21.7 | 42 | 7.6 | 8×5.5 |
| **FPC2** | 16 | 73.5 | 8.7 | 8.1×6.2 |
| **FPC3** | 16.2 | 108 | 3.5 | 9.4×4.1 |
| **FPC4** | 22 | 142 | 11 | 11.9×11.5 |
| **FPC5** | 48.9 | 114 | 9.2 | 7.0×1.5 |
| **FPC6** | 16.4 | 78.5 | 5.4 | 8.3× 4.8 |
| **FPC7** | 14.5 | 95.6 | 5.7 | 7.8×2.1 |
| **FPC8** | 21.2 | 94.2 | 4.6 | 6.8×5.5 |
| **FPC9** | 45 | 93.5 | 7 | 7.5×4.5 |
| **FPC10** | 27.2 | 89 | 2.9 | 6.5×3.7 |
| **FPC11** | 18.7 | 64 | 8.4 | 10.5×5.2 |

**Table S1** Dimensions of all FGT/Graphite/FGT devices.

### 3. Transmission Electron Microscopy (TEM) on FGT/Graphite/FGT heterostructures

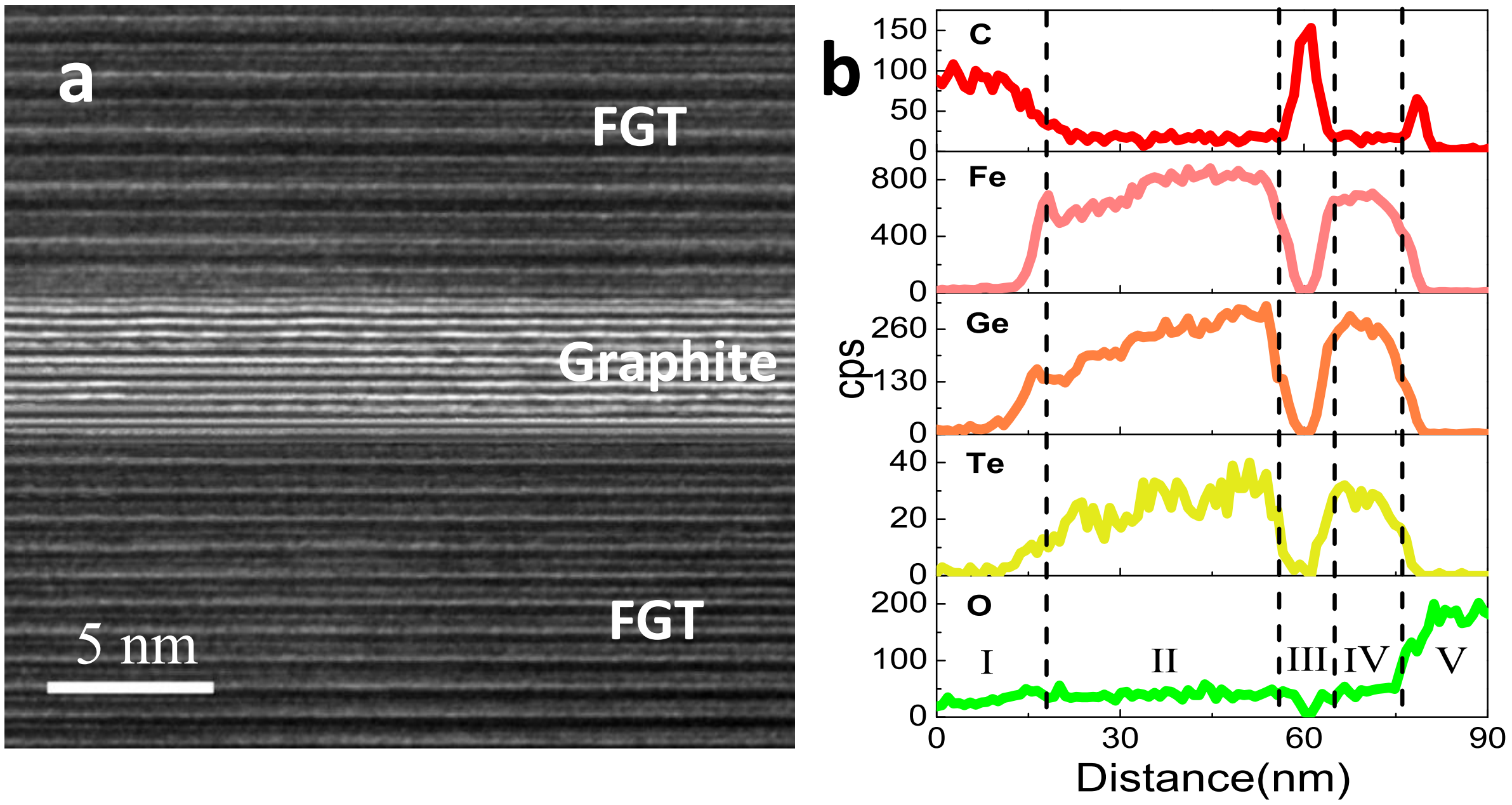

**Fig. S2 TEM on FGT/Graphite/FGT heterostructure. (a)** The cross-sectional TEM image obtained from a FGT/Graphite/FGT heterostructure. The TEM sample was prepared by FIB based ion milling. Interfaces between the graphite layer and the FGT layers are clearly observed. **(b)** EDX line-scan of the cross-sectional TEM image. Section I is the top protective layer composed of carbon. Sections II & IV are the FGT layers. Section III is the graphite layer. Section V is the $SiO_x$ (substrate) layer. The oxygen content is lower in the graphite layer and higher in FGT layers because the TEM sample was necessarily exposed to ambient and the graphite layer oxidizes less readily than the FGT.

## 4. The effect of graphite layer etch and samples with weak interlayer coupling

Etching the graphite is very important. As the resistivity values of the FGT and graphite are ~$10^{-8}$ $\Omega\cdot cm^{-1}$ and ~$10^{-4}$ $\Omega\cdot cm^{-1}$, respectively, current flow will occur mostly through the graphite layer if the horizontal size of the graphite layer is larger than the FGT layers.

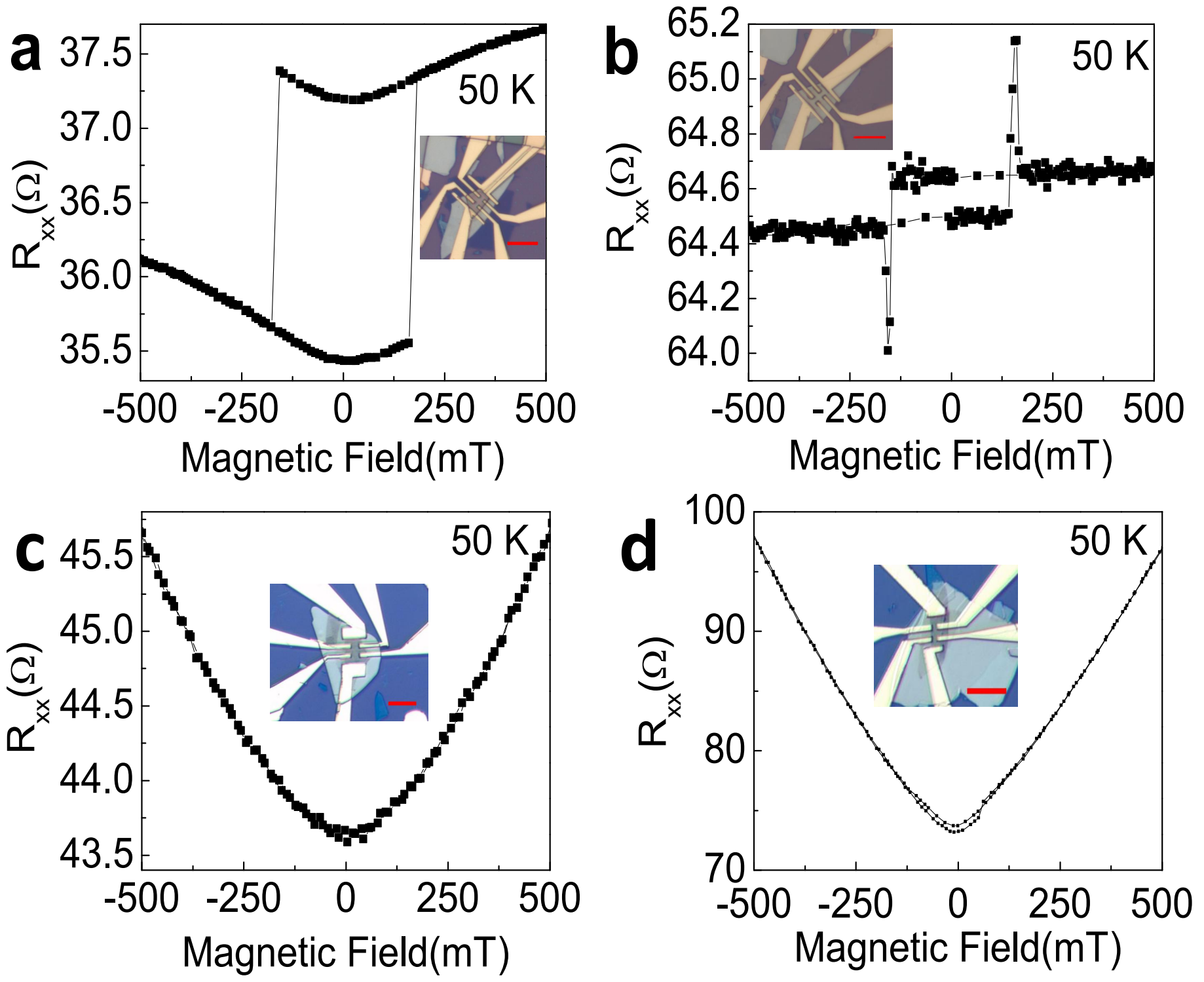

**Fig. S3 The effect of graphite layer etch and samples with weak interlayer coupling (a)** $R_{xx}(B)$ curve for sample FPC6 before the graphite layer was etched. The magnetoresistance of the graphite is very large, indicating that most of the electrons pass through the graphite layer directly. In this condition, the antisymmetric MR effect cannot be observed. Inset shows the device covered with a PMMA masking layer prior to $O_2$ etching. The scale bar represents 10 μm. **(b)** $R_{xx}(B)$ curve from sample FPC6 after the graphite layer was etched. An Ar plasma was used to etch the graphite layer after the metal deposition. Before the etch, a PMMA etching mask is fabricated by e-beam lithography. The electrons mainly pass through the tri-layer region and thus the antisymmetric MR effect can be displayed. Inset scale bar represents 10 μm. (**c and d**) Field dependent $R_{xx}$ of another 2 samples with etched graphite layer but contaminated interfaces between the graphite layers and the FGT layers. These samples only exhibit magnetoresistance from graphite. The inset red scale bars represent 10 μm. As shown in our experiments, a clean FGT/graphite interface is essential for the antisymmetric MR effect, which is indicative of the importance of the magnetic proximity effect. A short-ranged magnetic coupling at the FGT/graphite interface and/or induced magnetic order in the graphite layer may also affect the antisymmetric MR effect in addition to the spin-polarized 2DEG at the interface.

## 5. Fabrication of a symmetric Hall bar device based on FIB milling

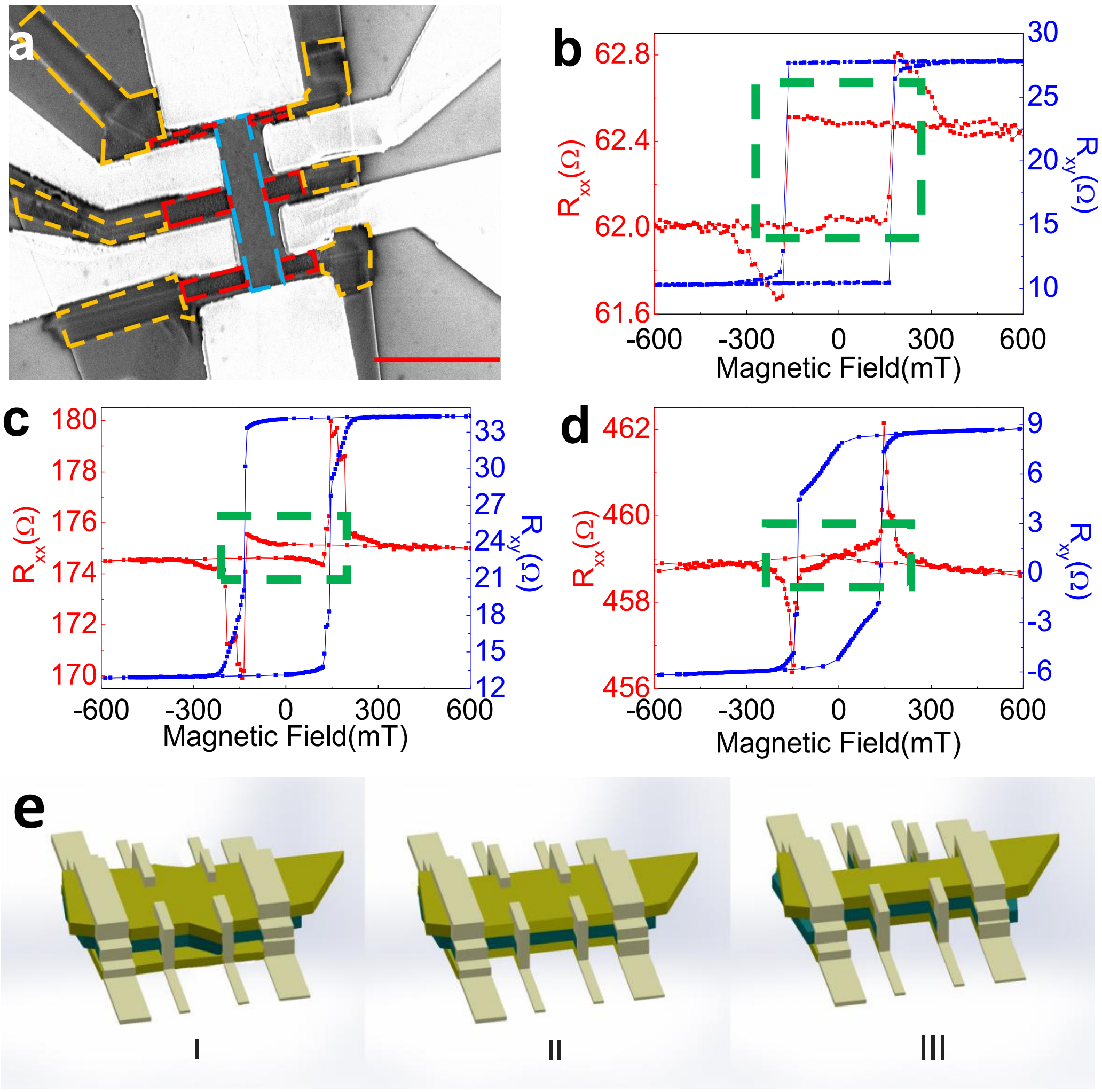

**Fig. S4 $R_{xx}$ and $R_{xy}$ of an FIB-etched FGT/graphite/FGT device (sample FPC5) at 50 K.** (**a**) Scanning electron microscope image of the FIB-etched device. The orange, red and blue dashed lines identify the etched region in the first etching step, the etched region in the second etching step, and the final un-etched region, respectively. The un-etched region remained non-exposed to the FIB. The scale bar represents 5 μm. (**b**) Field dependent $R_{xx}$ and $R_{xy}$ curves for the sample before the FIB etch process. At this step, only the graphite layer was etched by $O_2$ plasma (The blue layer in the Fig. S4e). (**c**) Field dependent $R_{xx}$ and $R_{xy}$ curves from the sample after the first etch. At this step, the heterostructure was etched into a symmetric shape, but not a strict Hall bar structure. (**d**) Field dependent $R_{xx}$ and $R_{xy}$ curves measured from the sample after the second etch. The device exhibited a standard Hall bar structure at this stage. From figure **b** to **d**, the "loops" in the green dashed line box gradually close, indicating a 3-state MR effect. (**e**) Schematic diagrams of each etch step. I, II, III representing **b**, **c**, **d**, respectively.

## 6. Transport measurement for other samples

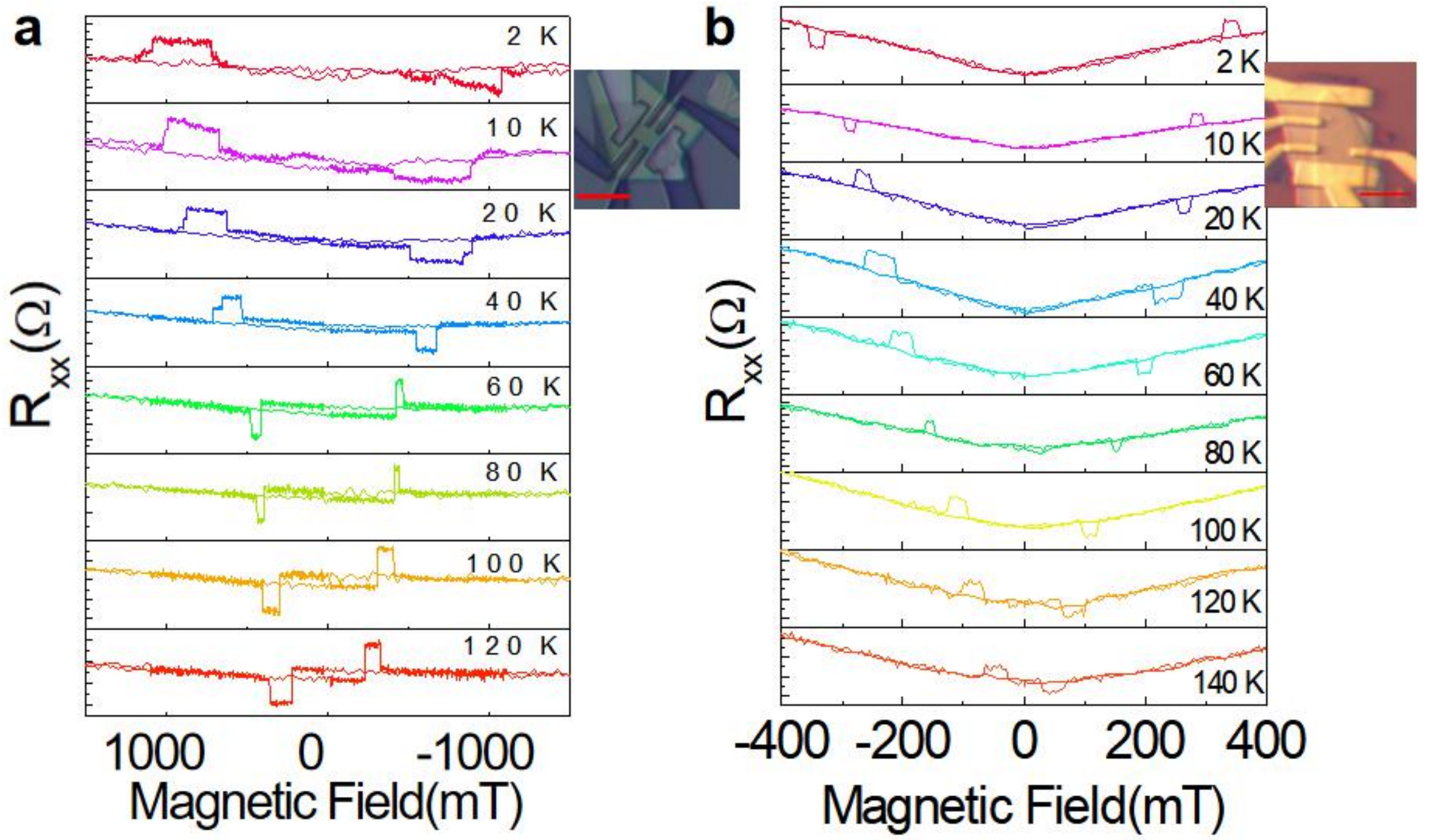

**Fig. S5 Measurements for FPC 1 and FPC9. (a)** $R_{xx}(B)$ curves at various temperatures measured from sample FPC1 tilted at 85°. This data was obtained from the same sample providing the data in figure 2 and figure 3 in the main text. The high resistance state and low resistance state switched when the temperature decreased from 60 K to 40 K due to the coercive field evolution of the top and bottom FGT layers. Some plateaus in the data obtained below 60 K also show multi-step behaviour. **(b)** $R_{xx}(B)$ curves from the sample FPC9 shown in the inset at different temperatures and with the magnetic field perpendicular to the sample surface. The high resistance state and low resistance state also switch between 20 K and 10 K. The magnitudes of the peaks remain similar at different temperatures. The scale bars represent 10 μm. Top FGT thickness: 45 nm, bottom FGT thickness: 93.5 nm, graphite thickness: 7 nm.

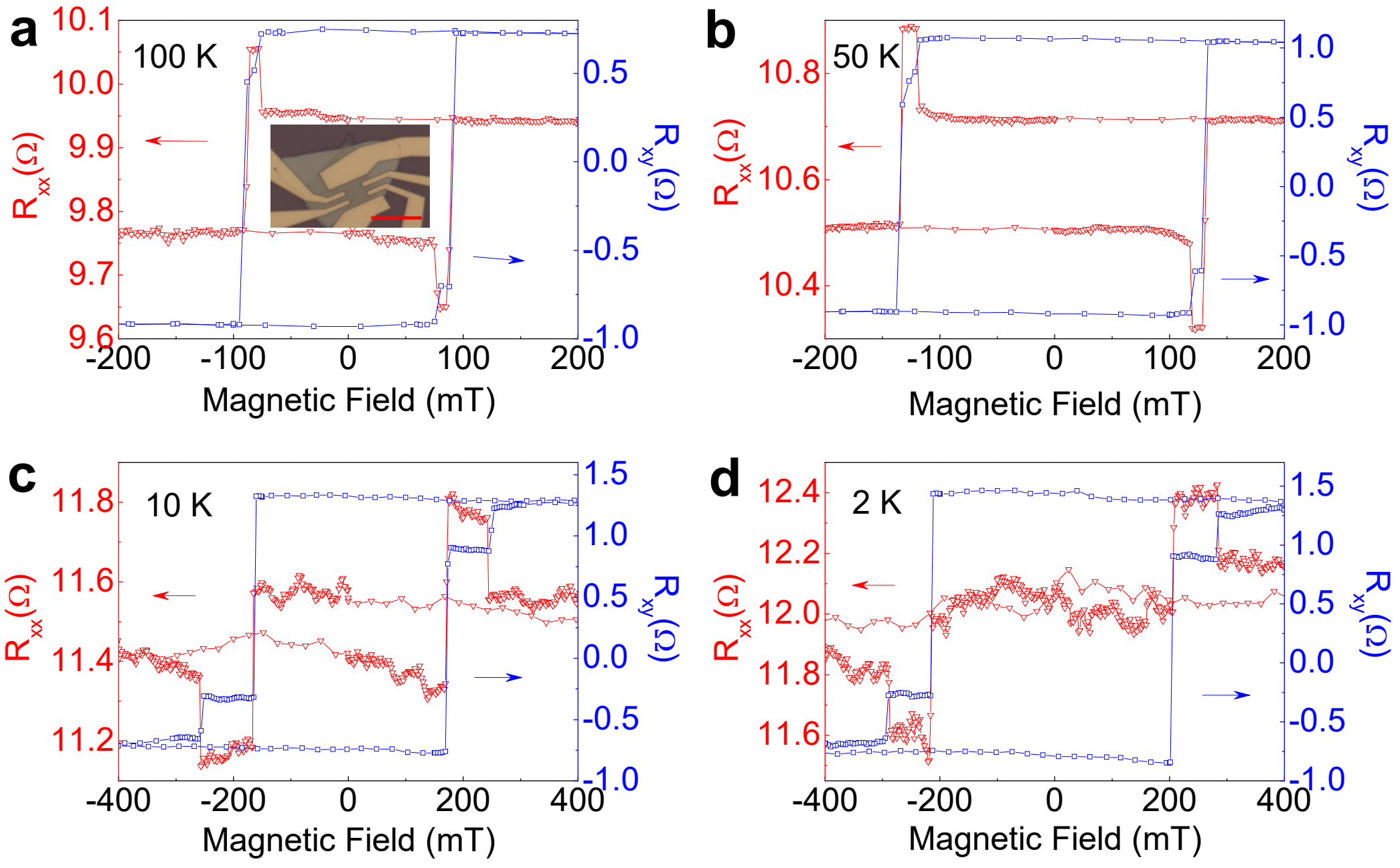

**Fig. S6 Measurement from an asymmetric sample FPC2 showing large anomalous Hall signal. (a)(b)** $R_{xx}(B)$ and $R_{xy}(B)$ curves measured from the sample at 100 K and 50 K. The contribution from the anomalous Hall effect is large in the $R_{xx}$ measurements. Inset of **(a)** shows an optical image of the device, the scale bar represents 10 μm. Top FGT thickness: 16.2 nm, bottom FGT thickness: 73.5 nm, graphite thickness: 8.7 nm. **(c)(d)** $R_{xx}(B)$ and $R_{xy}(B)$ curves measured from the sample at 10 K and 2 K. The direction of the plateaus in $R_{xx}$ switched in comparison with that observed at 50 K and 100 K, indicating different temperature dependence of the coercivity in the two FGT layers.

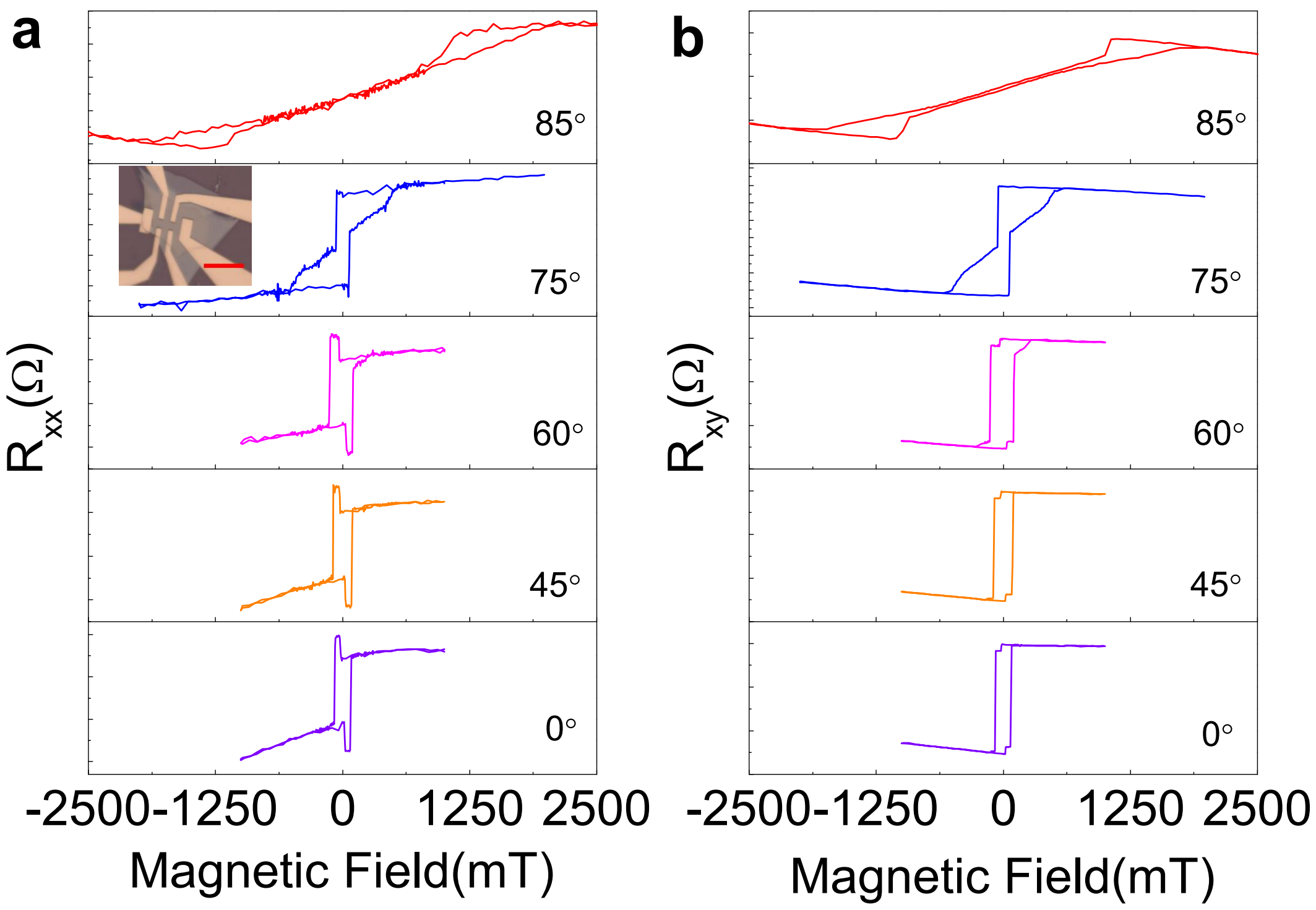

**Fig. S7 Angle dependent curves measured at 50 K for sample FPC11 with top and bottom FGT layers contacting each other. (a)** $R_{xx}(B)$ curves for the sample at various angles. When the angle > 60°, the plateaus nearly fade away and the shapes of $R_{xx}(B)$ curves are nearly the same as the corresponding $R_{xy}(B)$ curves. This is because the domain of the top and bottom FGT layer are coupled together and flip at the same time when the angle > 60°. **(b)** Corresponding $R_{xy}(B)$ curves. Inset of **(a)** shows an optical image of the device with scale bar representing 10 μm. Top FGT thickness: 18.7 nm, bottom FGT thickness: 64 nm, graphite thickness: 8.4 nm.

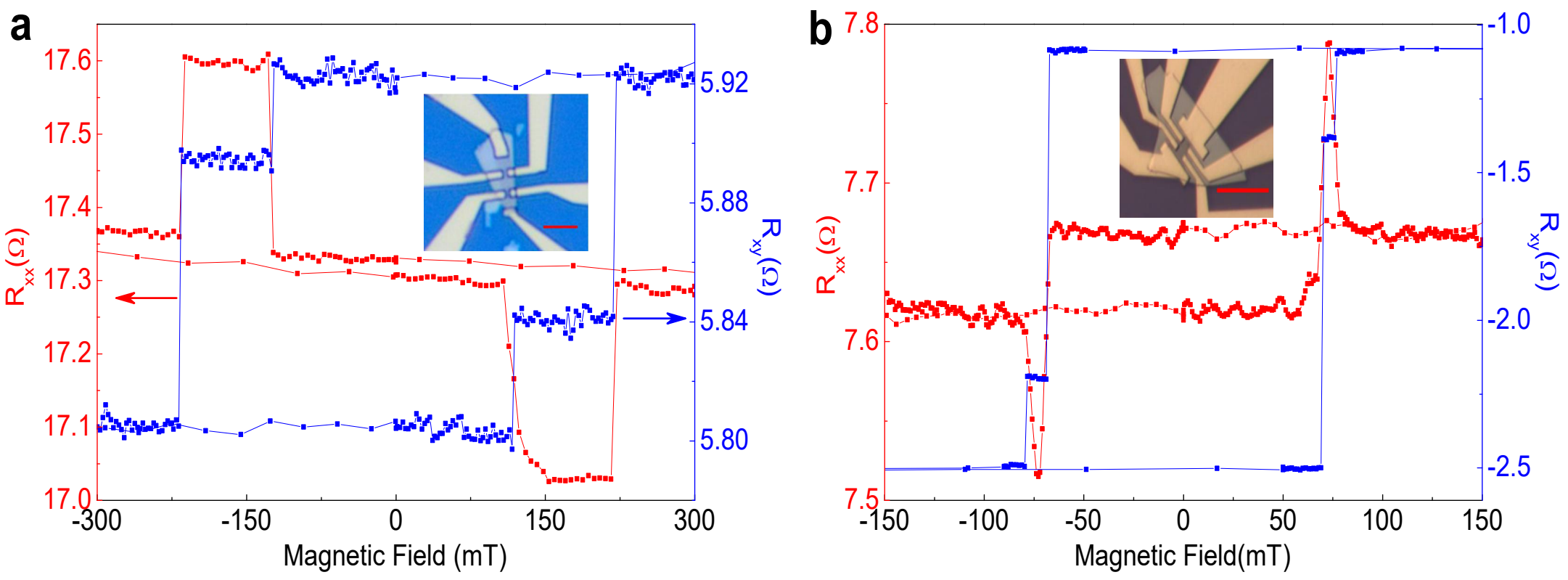

**Fig. S8 $R_{xx}$(B) and $R_{xy}$(B) curves for two samples with relatively larger $\Delta R_{xx}/R_{xx}$ value at 50 K. (a)** Sample FPC7, $\Delta R_{xx}/R_{xx}$ value is ~1.5%. Inset is the optical image for the device, the scale bar represents 5 μm. Top FGT thickness: 14.5 nm, bottom FGT thickness: 95.6 nm, graphite thickness: 5.7 nm. **(b)** Sample FPC8 exhibiting $\Delta R_{xx}/R_{xx}$ value of ~1.4%. The scale bar in the inset optical image represents 10 μm. Top layer thickness: 21.2 nm, bottom layer thickness: 94.2 nm, graphite thickness: 4.6 nm

## 7. Tentative resistor model

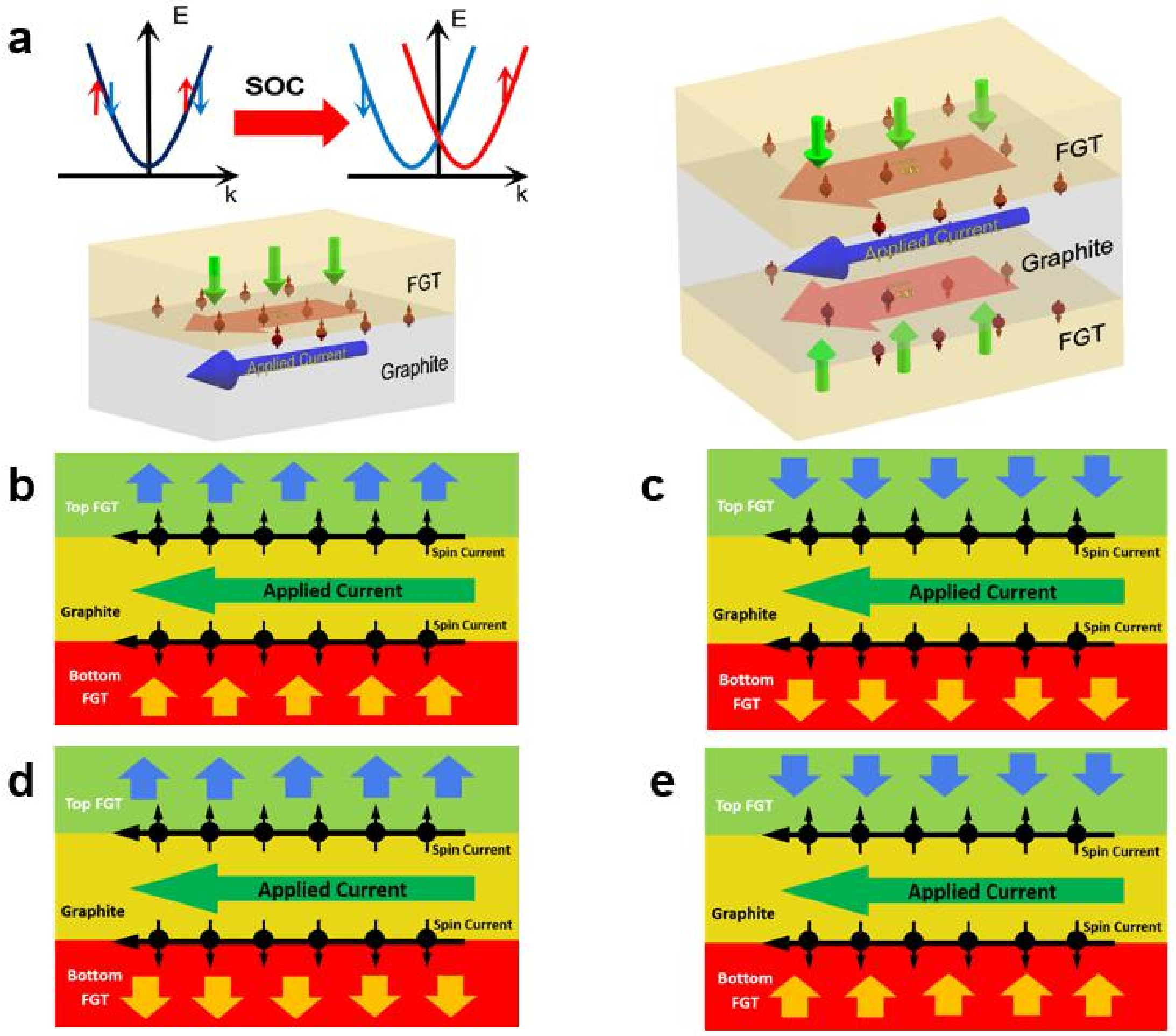

**Fig. S9 Tentative resistor model. (a)** Schematic diagram of the Rashba spin splitting induced spin distribution in the heterostructure. The green arrows represent the direction of magnetisation in the FGT. (**b and c**) The parallel magnetisation in a FGT/graphite/FGT heterostructure. The devices show intermediate interfacial conductance. (**d and e**) The antiparallel magnetisation in a FGT/graphite/FGT heterostructure. The devices show low and high interfacial conductance, respectively.

## 8. Band structure calculation

### 8.1 Computational methods

In this work, the Vienna Ab-initio Simulation Package (VASP) software package was used for first-principles calculations based on density functional theory (DFT) with the projector augmented wave method. The Predew-Burke-Ernzerhof (PBE) type generalized gradient approximation (GGA) was chosen to describe the exchange-correlation functional. For the electronic properties of bulk $Fe_3GeTe_2$, an energy cutoff of 310 eV was used for the plane wavefunction basis and the Brillouin Zone was sampled using a $7 \times 7 \times 2$ k-point mesh following the Monkhorst-Pack scheme. The surface states and surface spin texture were calculated using the WannierTools package and the tight-binding model Hamiltonian was derived from the Wannier functions.

### 8.2 Results and discussion

Bulk $Fe_3GeTe_2$ hosts a layered structure stacked along the [001] direction and the (001) surface can be regarded as a hexagonal lattice similar to graphene. We calculated the band structures with and without the SOC effect for the ferromagnetic $Fe_3GeTe_2$.

In the absence of SOC, the bands are decoupled into two spin channels. The Fermi level is buried deep in the bands in both spin channels, reflecting their metallic characteristics. After introducing the SOC effect, the bands split and a small gap opens. The projected band structures on the (001) surface are shown in Fig. S10(a), which is plotted along the k-path depicted in Fig. S10(b).

To study the surface spin behaviors of $Fe_3GeTe_2$ under the influence of intrinsic magnetic moment, we also calculated the surface Fermi arc and surface spin texture of $Fe_3GeTe_2$ with SOC. As illustrated in Fig. S10(b), distinct metallic features of both bulk and surface bands lead to the complex images of Fermi surface and surface Fermi arc, while the magnetization is mostly relaxed in the [001] direction.

By calculating the spin texture of surface states near the high symmetry points of Γ and K in the two-dimension Brillouin zone, we found that the spin is highly polarized along the z axis. The spin polarization directions can go up or down along the z axis for each band in the surface. For example, clear surface states with downward direction of spin polarization near Γ point are observed, as shown in Fig. S10(c). Near K point ( -K), the Fermi arcs form two loops centered on K (-K) point and exhibit similar spin polarization with Γ point, as shown in Fig. S10(d). Except for the clear surface states near the k-points Γ and K, some surface states within the bulk bands are found to exhibit upward spin polarization direction. We also found that, the spin direction will be reversed if the magnetization direction is reversed.

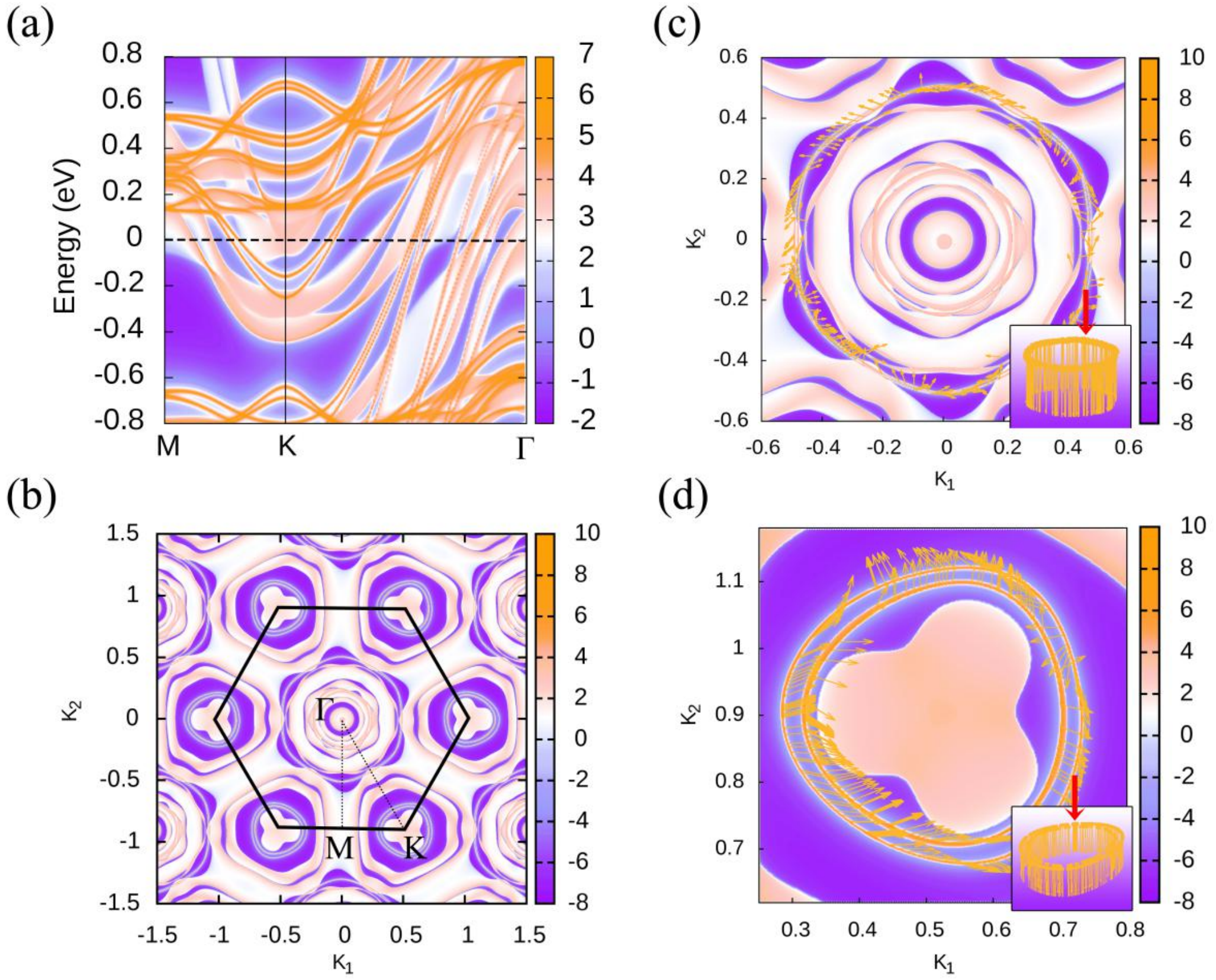

**Fig. S10 Surface states and surface spin texture of $Fe_3GeTe_2$. (a)** Surface states projected on the (001) surface along the k-path M-K-Γ as depicted in **(b)**. **(b)** Fermi arcs projected on the (001) surface and the first Brillouin zone of the slab systems. **(c)** Fermi arcs and surface spin texture around the Γ. **(d)** Fermi arcs and surface spin texture around the K point. The spin texture of the surface states in three dimension momentum space are shown in the insets.

**9. Discussion about the angle-dependent results in Fig. 3**

To obtain the antisymmetric MR effect in the FGT/graphite/FGT heterostructures, top and bottom FGT layers with different coercivities are required. As the coercivities of the FGT nanoflakes are related to their thicknesses, the antisymmetric MR is readily observed in heterostructures composed of FGT nanoflakes with different thicknesses.

We used a pick-up technique to fabricate the devices. In this technique, we first use the top FGT layer to pick up the graphite layer and then use the top FGT/graphite structure to pick up the bottom FGT layer. Finally, we release the heterostructures onto the $Si/SiO_x$ substrate. We prefer to choose thin FGT flakes as the top layer because a thin FGT nanoflake is transparent. In the stacking process, we can see through the top thin FGT layer and align the graphite layer (also thin and transparent) and the bottom FGT layer more easily.

It is well known that the coercivity of a ferromagnet is strongly related to the microstructure of the magnet. For a single crystalline FGT nanoflake, as shown in our previous work [*Nat. Commun.* 9, 1554 (2018)], the coercivity is mainly determined by the perpendicular anisotropy. However, the Fe stoichiometry, the thickness of the flake, the defects in the flake, the shape of the flake and the surface oxidation on the top FGT layer all contribute to a change in the coercivity.

When θ (the angle between the magnetic field and the perpendicular direction to the device surface) is small, the magnetic moments in an FGT nanoflake rotate coherently and show a single domain behavior (except in the regime near the coercivity), therefore the coercivity is mainly determined by the perpendicular anisotropic energy ($K_A$). If the $K_A$ value of the thinner FGT layer is smaller than that of the thicker FGT layer, the thinner layer will have a smaller coercivity and flip first at a lower field. When the θ value increases, the magnetic moments in the FGT will tilt towards the direction parallel to the surface and the defects near the surface will pin the rotation of the magnetic moments. This effect is larger in thinner FGT flakes, because it is a surface effect, and therefore can generate the gradual evolution of the coercivity as shown in Fig. 3 in the main text. When θ is large, the rotation of magnetic moments becomes non-collinear, which also generates different coercivity changes according to the microstructure (shape, defects, surface oxidation etc.) of the flake.

In all, the evolution of the coercivity with tilt angle is determined by the complex microstructure of the FGT nanoflakes. The device in Fig 3 shows the MR flips at $\theta = 70^{o}$ but the same effect may not appear in another device until the tilt angle is $90^{o}$. Nevertheless, all these magnetic moment configurations can be confirmed from the anomalous Hall measurements. No matter how the relative coercivity of the top FGT and bottom FGT changes, the “IN” and “OUT” magnetic moment configuration determines “high” or “low” MR states.